\documentclass[final,5p,times,twocolumn]{elsarticle}

\usepackage{amssymb}

\usepackage{xcolor}
\usepackage{listings}
\usepackage{subfig}
\usepackage{url}

\definecolor{lightgray}{gray}{0.9}
\definecolor{shadecolor}{gray}{0.92}
\definecolor{dkgreen}{rgb}{0,0.6,0}
\definecolor{gray}{rgb}{0.5,0.5,0.5}
\definecolor{mauve}{rgb}{0.58,0,0.82}

\journal{Future Generation Computer Systems}

\begin{document}

\begin{frontmatter}



\title{OMP4Py: a pure Python implementation of OpenMP}
\tnotetext[t1]{This work was supported by MICINN [PLEC2021-007662, PID2022-137061OB-C22]; Xunta de Galicia [ED431G 2019/04, ED431F 2020/08, ED431C 2022/16]; and European Regional Development Fund (ERDF).}


\author[1,2]{César Piñeiro\corref{cor1}}
\ead{cesaralfredo.pineiro@usc.es}
\author[1,2]{Juan C. Pichel}
\ead{juancarlos.pichel@usc.es}
\address[1]{Dept. of Electronics and Computer Science, Universidade de Santiago de Compostela, 15782 Santiago de Compostela, Spain}
\address[2]{Centro Singular de Investigaci\unexpanded{ó}n en Tecnolox\unexpanded{í}as Intelixentes (CiTIUS), Universidade de Santiago de Compostela, 15782 Santiago de Compostela, Spain}
\cortext[cor1]{Corresponding author}

\begin{abstract}
Python demonstrates lower performance in comparison to traditional high performance computing (HPC) languages such as C, C++, and Fortran. This performance gap is largely due to Python's interpreted nature and the Global Interpreter Lock (GIL), which hampers multithreading efficiency. However, the latest version of Python includes the necessary changes to make the interpreter thread-safe, allowing Python code to run without the GIL. This important update will enable users to fully exploit multithreading parallelism in Python. In order to facilitate that task, this paper introduces OMP4Py, the first pure Python implementation of OpenMP. We demonstrate that it is possible to bring OpenMP's familiar directive-based parallelization paradigm to Python, allowing developers to write parallel code with the same level of control and flexibility as in C, C++, or Fortran. The experimental evaluation shows that OMP4Py significantly impacts the performance of various types of applications, although the current threading limitations of Python’s interpreter (v3.13) reduce its effectiveness for numerical applications.
\end{abstract}




\end{frontmatter}


\section{Introduction}

Lately, Python has become the most popular programming language~\cite{Tiobe}. Its ease of use has led to its widespread adoption across many scientific domains, from data analysis to machine learning. Despite its advantages in these areas, Python significantly lags behind traditional low-level HPC (High Performance Computing) languages like C/C++ and Fortran when it comes to achieving high performance. This performance gap can be attributed to two primary factors. First, Python's interpreted nature introduces overhead that impacts execution speed compared to the compiled code of languages such as C and Fortran. Second, Python's Global Interpreter Lock (GIL) limits its ability to fully exploit multithreading, hindering its scalability in parallel computing tasks. These limitations present challenges to utilizing Python in HPC environments, where speed and efficiency are crucial.

Various efforts have been made to improve support for multithreading parallelism in Python, but important limitations remain~\cite{PEP703}. However, a more definite solution is expected with the latest Python interpreter, whose final release was in October 2024. This version includes the necessary changes to make the interpreter thread-safe, allowing Python code to run without the GIL.

Based on this important advancement, this paper introduces OMP4Py\footnote{It is publicly available at \url{https://github.com/citiususc/omp4py}}, the first native Python implementation of OpenMP~\cite{padua2011}, a widely recognized standard programming model for exploiting multithreading parallelism in HPC. OMP4Py integrates OpenMP functionalities into Python through two main mechanisms. First, it adapts OpenMP's directive-based approach, allowing Python users to embed parallel constructs directly into their source code. These transformer directives enable parallel execution by instructing OMP4Py to modify the Python code accordingly. Second, OMP4Py includes a set of runtime library functions that closely mirror those provided by OpenMP. These functions allow users to manage parallel execution parameters, such as the number of threads and scheduling policies, giving them flexibility to fine-tune the parallel behavior of their Python programs. In this way, we prove that OpenMP's directive-based parallelization model can be seamlessly integrated into Python, giving developers the same level of control in writing parallel code as they would have in C, C++, or Fortran.

A thorough experimental evaluation was carried out in the paper, demonstrating that OMP4Py has significant potential for hybrid parallelism in combination with mpi4py~\cite{dalcin21} and for non-numerical workloads. However, the current threading limitations of Python's interpreter (v3.13) hinder its scalability for numerical applications. It is important to highlight that as these limitations in the Python interpreter are progressively resolved, the scalability constraints of OMP4Py when running numerical applications will gradually disappear, without requiring any modifications to its implementation.

The paper is structured as follows: Section \ref{sec:background} provides some background and summarizes previous research in the field. Section \ref{sec:omp4py} introduces OMP4Py, detailing its design, features, and implementation. Section \ref{sec:exp_results} presents the experimental results, discussing various experiments conducted to evaluate OMP4Py’s performance across different types of applications. Finally, Section \ref{sec:conclusions} summarizes the key findings and proposes directions for future research..

\section{Background \& Related work}
\label{sec:background}

\subsection{OpenMP}
\label{sec:openmp}

OpenMP~\cite{padua2011} is a parallel programming model originally designed for shared-memory computer systems, aiming to simplify the exploitation of inherent concurrency in many algorithms. OpenMP primarily follows the fork-join model of parallel execution. In this model, the program starts with a single \emph{initial thread}. At parallel regions, the initial thread creates multiple parallel threads that concurrently execute the assigned tasks. Once the parallel tasks are completed, the threads join back into the initial thread, which continues executing the program sequentially. Through compiler directives, programmers can instruct the compiler to generate multithreaded code at a higher level of abstraction, avoiding the manual management of thread creation and task assignment required by low-level approaches like pthreads in the POSIX library. The OpenMP API is compatible with C/C++ and Fortran.

Although initially designed for shared-memory architectures, OpenMP has expanded its capabilities to support heterogeneous computing. Since the introduction of the \texttt{target} directive family, OpenMP has enabled offloading computations to accelerators, such as GPUs, which often employ distributed-memory models internally. This extension allows OpenMP to be used in hybrid computing environments where both shared- and distributed-memory paradigms coexist. For this reason, OpenMP remains the standard for exploiting multithreading capabilities of modern multi-core CPUs while also enabling high-performance execution on heterogeneous architectures.

The OpenMP API standard specification\footnote{\url{https://www.openmp.org/specifications} [online, accessed November 5, 2024]} started in 1997 (version 1.0), and it continues to evolve, with new constructs and features being added over time. The latest release, version 6.0, was recently published in November 2024. However, most OpenMP programmers typically use only a subset of the OpenMP 3.0 specification released in 2008. This subset, referred to as the \emph{OpenMP Common Core}~\cite{mattson2019}, comprises the 21 most commonly utilized elements of OpenMP.  Here are some of the main constructs, clauses and functions included in the OpenMP Common Core:
\begin{enumerate}
    \item Parallel regions:
    \begin{itemize}
    \item \texttt{\#pragma omp parallel} - Defines a parallel region, which is a block of code executed by a team of multiple threads. According to OpenMP terminology, a \emph{primary thread} is the thread that encounters a parallel construct, creates a team of threads, generates a set of implicit tasks, and executes one of those tasks as thread number 0. If there is only a single team of threads, the initial thread and the primary thread refer to the same thread.
    \end{itemize}
    \item Work sharing constructs:
    \begin{itemize}
    \item \texttt{\#pragma omp for} - Distributes loop iterations among threads in a parallel region. There also exists a combined construct equivalent to a \texttt{parallel} construct followed by a \texttt{for} (\texttt{\#pragma omp parallel for}).
    \item \texttt{\#pragma omp sections} - Divides work into separate sections that can be executed in parallel.
    \item \texttt{\#pragma omp single} - Specifies a block of code that should be executed by only one thread.
    \end{itemize}
    \item Tasking:
    \begin{itemize}
    \item \texttt{\#pragma omp task} - Creates an explicit task for deferred execution within the construct.
    \item \texttt{\#pragma omp taskwait} - Ensures that all child tasks created within the current task are completed before the execution of the program continues.
    \end{itemize}
    \item Synchronization constructs:
    \begin{itemize}
    \item \texttt{\#pragma omp barrier} - All threads in the current team must reach a barrier before any can continue.
    \item \texttt{\#pragma omp critical} - Ensures that only one thread at a time executes a block of code.
    \end{itemize}
    \item Scheduling and other clauses:
    \begin{itemize}
    \item \texttt{schedule(static [,chunk])} - Distributes the iterations of a loop in contiguous blocks, with each thread receiving a block of iterations of size \texttt{chunk}.
    \item \texttt{schedule(dynamic [,chunk])} - Dynamically assigns chunks of iterations to threads at runtime.
    \item \texttt{nowait} -It is used to remove the implicit barrier at the end of certain constructs such as \texttt{for}, \texttt{sections} and \texttt{single}.
    \end{itemize}
    \item Data environment clauses:
    \begin{itemize}
    \item \texttt{private(list)} -  Specifies that each thread should have its own private copy of the variables in the list.
    \item \texttt{firstprivate(list)} - Similar to \texttt{private}, but the variable is initialized using the value from the initial thread.
    \item \texttt{lastprivate(list)} - Ensures that the value of a private variable from the last iteration is copied back to the original variable.
    \item \texttt{shared(list)} - Specifies that variables in the list should be shared among all threads in the current team.
    \item \texttt{reduction(op:list)} - Performs a reduction operation on variables across all threads in a team in a parallel region.
    \end{itemize}
    \item Functions:
    \begin{itemize}
    \item \texttt{omp\_set\_num\_threads(int)} - Sets the number of threads to be used in subsequent parallel regions, unless overridden by a more specific mechanism (e.g., the \texttt{num\_threads} clause in a parallel directive) or modified by another call to \texttt{omp\_set\_num\_threads}.
    \item \texttt{omp\_get\_thread\_num()} - Returns the unique thread number of the calling thread within its team.
    \item \texttt{omp\_get\_wtime()} - Returns the elapsed time since some point in the past. It is used to measure the execution time of a segment of code.
    \end{itemize}
\end{enumerate}

\subsection{Limitations of Python for multithreading parallelism}
\label{sec:limitations_python}

As commented previously, Python has emerged as the most popular programming language in recent years~\cite{Tiobe}, renowned for its simplicity, productivity, and readability. As a result, we can find Python applications in practically all scientific areas. However, when the main goal is obtaining high performance, Python falls far behind traditional low-level HPC languages such as C and Fortran. There are two main reasons for that behavior. 

First, Python is an interpreted programming language, which causes an important overhead due to the need for real-time translation of source code into machine code during runtime. This problem can be mitigated using Just-In-Time (JIT) compilers like Numba~\cite{Numba}, which translates Python functions to optimized machine code at runtime using the LLVM compiler. This approach aims to achieve performance levels comparable to those of C or Fortran. However, Numba is optimized for numerical computations, so code that involves extensive string manipulation, complex data structures, or I/O operations may not see significant performance gains. This is the case, for example, of purely Pythonic code that relies heavily on built-in data structures like lists and dictionaries.

The second issue is related to how Python handles multithreading due to the existence of the Global Interpreter Lock (GIL). The GIL is a locking mechanism that protects access to Python objects, preventing multiple threads from executing Python code simultaneously. The GIL was originally introduced to simplify thread management and protect against race conditions and memory corruption, making it easier for developers to write concurrent code safely. However, since the GIL allows only one thread to execute Python code at a time, multithreading in Python is unsuitable for CPU-bound tasks where the performance gain from parallel execution is significant. In this way, the GIL becomes the most important obstacle to take advantage of parallelism using multi-core CPUs efficiently in Python. Note that, on the other hand, I/O-bound tasks, such as network requests or file operations, can benefit from multithreading in Python. In these cases, the GIL is released when a thread performs I/O operations, allowing other threads to execute Python code.  

\subsection{Advances towards a multithreading-friendly Python}

Python currently offers several methods to enable parallelism, but these techniques have significant limitations~\cite{PEP703}. For instance, the multiprocessing library allows programs to create and interact with subprocesses. This facilitates parallelism because each subprocess has its own Python interpreter, meaning there is one GIL per process. Some examples of applications and libraries that use the multiprocessing package to parallelize tasks are the deep learning frameworks PyTorch~\cite{ansel2024} and TensorFlow~\cite{abadi2016}, and the HPC-Big Data framework IgnisHPC~\cite{pineiro2022}. However, multiprocessing has some important drawbacks. Communication between processes is limited, as objects typically need to be serialized or copied to shared memory. This introduces overhead and complicates building APIs on top of multiprocessing. In addition, starting a subprocess is more expensive than starting a thread. Finally, many C and C++ libraries support multithreading but do not support access or usage across multiple processes.

An additional solution to take advantage of parallelism in Python (multithreading in this case) is based on the observation that functions implemented in C may use multiple threads internally. For instance, Intel’s NumPy distribution, among others, employs this approach to internally parallelize individual operations. This works well when the basic operations are large enough to be parallelized efficiently, but not when there are many small operations or when the operations depend on some Python code. Note that invoking Python from C requires obtaining the GIL, which means even small pieces of Python code can prevent scaling.

Another interesting alternative to bring parallel multithreading to Python is PyOMP~\cite{anderson2021,mattson2021}, which is  a prototype system with support for OpenMP. As noted earlier, OpenMP API tells the compiler how to generate multithreaded code. In the case of PyOMP, Numba acts as the compiler, transforming Python code into LLVM, thereby bypassing the GIL. Note that the Numba compiler works with NumPy arrays, which must be used for any arrays inside a PyOMP function. However, as stated previously, Numba, and consequently PyOMP, struggle when calling functions or interacting with Python objects. Due to the lack of compiler directives in Python, PyOMP uses the \texttt{with} statement instead. In this way, for example, to create a parallel region uses: \texttt{with openmp("parallel"):}, while the equivalent directive in C/C++ API would be: \texttt{\#pragma omp parallel}. On the other hand, PyOMP supports 90\% of the OpenMP Common Core missing \texttt{nowait} and the \texttt{dynamic} schedule (see Section \ref{sec:openmp}).

As we have already discussed, the main factor that severely limits the concurrency of multithreaded Python code is the GIL. Despite several efforts over the years to remove the GIL~\cite{Gilectomy,PythonNoGIL}, none have been considered for inclusion in the Python interpreter until recently~\cite{PEP703}. As a result, Python 3.13, whose final release was in October 2024, includes a build configuration flag (\texttt{-{}-disable-gil}) to allow it to run Python code without the Global Interpreter Lock and with the necessary changes needed to make the interpreter thread-safe. This will initiate a long process toward making the disabling of the GIL the default option in the Python interpreter. Note that a basic JIT compiler was also added to Python 3.13~\cite{PEP744}, aiming to reduce the performance distance with compiled languages such as C and C++. 

\section{OMP4Py}
\label{sec:omp4py}

OMP4Py is a novel implementation of the OpenMP standard, designed specifically for Python. It currently supports the complete specifications of version 3.0. As explained in Section \ref{sec:openmp}, this ensures that we cover most of the needs of OpenMP HPC applications and programmers. This tool was developed with adherence to the official OpenMP documentation\footnote{\url{https://www.openmp.org/wp-content/uploads/spec30.pdf} [online, accessed November 5, 2024]}, which outlines the necessary compiler directives, runtime library functions, and environment variables required for creating shared memory parallel programs using threads. The OpenMP standard traditionally supports C, C++, and Fortran, which are low-level, compiled languages that use specific syntax for parallelism directives, as defined in the language standard. These directives are part of the OpenMP specification, and while they may differ in syntax between languages (such as \texttt{\#pragma} in C/C++ or sentinel \texttt{!\$} in Fortran), they all serve the same purpose: to manage parallelism during compilation.
The main goal of OMP4py is to bring the familiar parallelization paradigm of OpenMP to Python, allowing Python developers to write parallel code with the same level of control and flexibility as in C, C++, or Fortran. This tool aims to port the OpenMP model, with all code execution handled natively using Python threads. This ensures integration with Python’s libraries, enabling multi-threaded performance directly within Python's ecosystem.

OMP4Py integrates the core functionalities of OpenMP into Python in the following ways:

\begin{itemize}
    \item \emph{Transformer directives}: Adapting OpenMP’s directive-based approach, OMP4Py allows Python users to embed parallel constructs directly into their source code. These directives instruct the OMP4Py to transform the Python code for parallel execution.

    \item \emph{Runtime library functions}: OMP4Py includes a set of runtime library functions that mirror those provided by OpenMP. These functions manage parallel execution parameters, such as the number of threads and scheduling policies, among others. This feature provides users the flexibility to fine-tune the parallel behavior of their Python programs.
\end{itemize}

Since Python lacks a preprocessor, we needed to integrate OpenMP directives directly into the Python language while adhering to Python's best practices and idioms. To achieve this, we defined a function \texttt{omp} that operates similarly to OpenMP directives in C/C++, maintaining the same syntax and functionality. The function itself has no effect when executed; it serves solely as a container for the OpenMP directives. In this way, preprocessor directives in C/C++, such as:
\begin{center}
\texttt{\#pragma omp parallel num\_threads(2)} \\  
\end{center}
would be integrated into Python using OMP4Py as:
\begin{center}
\indent\texttt{with omp("parallel num\_threads(2)"):} \\
\end{center}
Note that when a OpenMP directive must be used within structured blocks, the \texttt{omp} function is used together as part of a \texttt{with} block (similar to PyOMP syntax~\cite{mattson2021}); otherwise, it is used as a standalone function call.
Finally, by themselves, calls to the \texttt{omp} function, as mentioned earlier, have no effect. Therefore, we must instruct the Python interpreter to restructure the code according to the content of each OpenMP directive before executing it. To accomplish this task, we employ Python \emph{decorators}. A decorator can be applied to a function or class to change its behaviour in a very elegant and intuitive way. So, we must decorate a function or class containing the OpenMP directives with the \texttt{@omp} decorator.

\begin{figure}[t!]
    \begin{lstlisting}[language=Python]
    from omp4py import *
    import random 
    
    @omp
    def pi(num_points):
        count = 0
        with omp("parallel for reduction(+:count)"):
            for i in range(num_points):
                x = random.random()
                y = random.random()
                if x * x + y * y <= 1.0:
                    count += 1
        pi = 4 * (count / num_points)
        return pi

    print(pi(10000000))    
    
    \end{lstlisting}
	\vspace{-0.35cm}
	\caption{Example of a Monte Carlo method for $\pi$ calculation using OMP4Py.}
	\label{fig:pi_calculation}    
\end{figure}

Figure \ref{fig:pi_calculation} shows an example of Python code for the parallel calculation of $\pi$ using OMP4Py. First, in line 4, the pi function is decorated with \texttt{@omp}, indicating that it contains OpenMP directives that need to be processed. Next, in line 7, a parallel region is started using \texttt{omp("parallel for reduction(+:count)")}. This statement instructs OMP4Py to parallelize the following for loop, where each thread contributes to the reduction operation on \texttt{count}. Finally, line 14 returns the computed value of $\pi$.

To achieve parallel code generation in Python, we must consider that Python is an interpreted language. Unlike lower-level languages defined in the OpenMP standard, there is no compilation process where we can apply the transformations needed to produce parallel code. Instead, following Python's philosophy, OpenMP code is generated by the interpreter itself at the moment the module containing the user's code is loaded. For a Python module to begin execution, all global definitions must be loaded, which can include the importation of other modules, global variables, and, relevant to our case, the declaration of functions or classes. When we apply a decorator to a function or class, the interpreter will invoke this decorator with the function or class as arguments right after it has finished loading it. The result of executing the decorator will replace the original function or class, and once the module is fully loaded, the user's code will only interact with the decorated version.

The \texttt{omp} decorator aims to replace all directives within the source code of a function or class and generate a new version with parallel behavior. The first step is to obtain information about the object representing the decorated function or class. For this, we use the \texttt{inspect} module, which allows us to retrieve the source code of the object. Once we have the source code, we generate an abstract syntax tree (AST) using Python's \texttt{ast} module. The AST provides a simple and transformable representation of the source code. The transformation process involves an in-order traversal of the AST. Each time a directive is encountered, it is parsed and checked for errors before applying the transformations to the AST. If any errors are found, the interpreter will abort with a \texttt{SyntaxError}, just as it would when encountering invalid syntax in Python code. After all directives have been processed, the resulting tree is transformed into object code using the \texttt{compile} function and loaded into the interpreter with \texttt{exec}. Finally, the decorator returns the new function or class with the parallel code, replacing the original.

\subsection{Parallel Directive}

In OpenMP, the \texttt{parallel} directive is essential for creating parallel programs. This construct allows a specific segment of code to execute simultaneously across multiple threads. When a code block is preceded by this construct, the program creates a team of threads, and each thread, including the original one, executes a copy of the code block concurrently. Multiple parallel directives can be nested, enabling each thread to create new teams of threads that work independently. This feature is known as \emph{nested parallelism} and must be enabled with \texttt{omp\_set\_nested}. By definition in the standard, every OpenMP program runs within an implicit single-threaded parallel construct. This allows API functions like \texttt{omp\_get\_num\_threads} or \texttt{omp\_get\_thread\_num}, among others, to operate independently of where they are called in the code. If multiple nested \texttt{parallel} directives are present and \texttt{omp\_set\_nested} is disabled, only the outermost directive spawns threads. Subsequent parallel regions do not create new threads; instead, only the existing threads that encounter the directive will execute the code.

Variables defined within a parallel block are local to each thread, meaning that each thread has its own copy. However, previously defined variables can be either \texttt{private} or \texttt{shared}, according to the user's preference. Shared variables maintain the same value across all threads in the current team. Any modifications made by one thread are visible to the others, and this shared value persists even after the parallel block ends.  By default, all variables defined before a parallel block are shared. While users can explicitly mark variables as shared, they only need to specify which variables should be private. Declaring a variable \texttt{private} makes it behave as if it were created inside the parallel block, with each thread starting with an uninitialized value. Within the parallel region, the private variable is treated independently by each thread, and its final value is discarded after the block ends. The outer variable remains unaffected by any changes made to the private copy during the execution. To retain the value of the outer variable, users can employ \texttt{firstprivate}, which initializes each thread's local copy to the variable's previous value.

\begin{figure}[t!]
    \begin{lstlisting}[language=Python]
from omp4py import *

@omp
def f(a):
    b = "1"
    c = -1
    d = [1, 2]
    e = True
    with omp("parallel shared(b) private(c) firstprivate(d) num_threads(4)"):
        f = omp_get_thread_num()
        a = 1
        c = f
        d.append(3)
        print(b, c, d, f)
            
    print(a, c)    
    \end{lstlisting}
    
    \begin{lstlisting}[language=Python]
from omp4py import *

def f(a):
    b = "1"
    c = -1
    d = [1, 2]
    e = True
    def _omp_parallel_1():
        nonlocal a, b
        _omp_c_2 = None
        _omp_d_3 = _omp_copy(d)
        f = omp_get_thread_num()
        a = 1
        _omp_c_2 = f
        _omp_d_3.append(f)
        print(b, _omp_c_2, _omp_d_3, f)
 
    _omp_parallel_run(_omp_parallel_1, num_threads=4)
            
    print(a, c) 
    \end{lstlisting}
	\vspace{-0.35cm}
	\caption{Example of the \texttt{parallel} directive: user code (top) and its corresponding translation by OMP4Py (bottom).}
	\label{fig:omp_parallel}    
\end{figure}

The implementation of the \texttt{parallel} directive in Python presents two main challenges. First, the code that is meant to run in parallel across different threads must be placed inside a function. Second, it is necessary to manage variables used both inside and outside the code block to determine whether they should be shared or private for each thread. Figure \ref{fig:omp_parallel} shows an example of a user-defined parallel block (top) and the corresponding code generated by OMP4Py for its parallel execution (bottom). It is important to note that, in code generation examples, the prefix \texttt{\_omp\_} will be used for all internal OMP4Py symbols, with dynamically created symbols including a number to avoid collisions. The example shows a simple function with a parallel directive and six representative variables ('\texttt{a}' to '\texttt{f}') with different types. The first change to observe is that the decorator (line 3, top code) and the function containing the directive (line 9, top code) have been removed. Once the code is transformed, the decorator is no longer needed and must be removed to prevent multiple processing. The parallel directive (line 9, top code) has been transformed into a function (line 8, bottom code) that will be called by different threads. The \texttt{\_omp\_parallel\_run} function (line 18, bottom code) is an internal OMP4Py function responsible for initializing and launching threads using Python's Thread class. Since the user has specified a number of threads with \texttt{num\_threads}, we also need to pass this argument. Note that the existing thread, or initial thread, is an execution thread as well, so only $num\_threads - 1$ additional threads are created.

The next step is to deal with local variables. The parallel block (lines 9-16, bottom code) is now within a nested function, a term for functions that are defined inside another function. In Python, nested functions can access variables from the outer function. First, variables \texttt{a} and \texttt{b} are shared; in this case, we implicitly use the \texttt{nonlocal} keyword (line 9, bottom code) to indicate that any new assignment modifies the variable in the outer function rather than creating a new one inside the nested function. Additionally, variables \texttt{c} and \texttt{d} are declared private, so we create two new variables and replace all their uses within parallel block code. Variable \texttt{c} has no initial value (line 10, bottom code), while \texttt{d} is initialized using \texttt{\_omp\_copy}, which creates a shadow copy of the variable (line 11, bottom code). Finally, variable \texttt{e} is not used within the code, and \texttt{f} is a local variable, so they are ignored. Consequently, we observe that the internal print statement (line 16, bottom code) will show the same shared value of \texttt{b} in each thread, while the other values will differ for each thread. However, the external print statement (line 20, bottom code) will display the modified value of \texttt{a} from within the parallel block (line 13, bottom code) but will retain the initial value of \texttt{c} (line 5, bottom code).

\subsection{Worksharing Constructs}

The distribution of work among threads is the base of OpenMP's functionality. A worksharing construct divides execution regions among the team of threads created in the most recent parallel directive. Common worksharing constructs include \texttt{for} loops, which distribute loop iterations among threads, \texttt{sections}, which divide distinct code blocks among threads, and \texttt{single}, which ensures that a specific section of code is executed by only one thread. 
Worksharing regions are executed as soon as the first thread encounters them. While the first thread can begin executing its part of the work immediately, it must remain in the worksharing region until all threads in the team reach the same point. This is the default behavior for all worksharing constructs, although users can modify it if necessary using the \texttt{nowait} clause. If there is only a single thread, the work is still distributed according to the scheduling policy, although in this case, the work will be executed sequentially since only one thread is available to process it.

\subsubsection{For}
\label{sec:for_construct}

The \texttt{for} directive is used to parallelize loops, allowing iterations to be distributed across multiple threads. Iterations are grouped into chunks and assigned to each thread according to a scheduling policy. OpenMP defines three scheduling policies: \emph{static}, \emph{dynamic}, and \emph{guided}. The static policy assigns chunks to threads in a round-robin fashion, the dynamic policy assigns chunks as they are requested by threads using a shared variable index, and the guided policy is similar to dynamic but the chunk size decreases dynamically with each assignment. Additionally, \emph{auto} allows the compiler to choose the scheduling policy, while \emph{runtime} defers the decision to runtime, where it is determined by an environment variable.

The first challenge to implement the \texttt{for} directive in Python is the absence of the traditional \texttt{for} loop. In Python, all \texttt{for} loops are constructed as \texttt{foreach} loops that iterate over a collection of elements. Consequently, to achieve the equivalent of a classic \texttt{for} loop, one must use a \texttt{foreach} loop over a range of values generated by the built-in \texttt{range} function. For example, a simple C-style loop like \texttt{for(int i = 0; i < 10; i++)} is equivalent to \texttt{for i in range(0, 10, 1)} in Python, where the initial value, final value, and step are specified as arguments. It is important to note that ranges cannot be used directly to divide the work because Python iterators are sequential, while C++ iterators allow random access, enabling direct access to any element, size calculation, and arithmetic operations between iterators.

\begin{figure}[t!]
    \begin{lstlisting}[language=Python]
from omp4py import *

@omp
def f():
    xs = [0] * 20
    with omp("parallel"):
        with omp("for schedule(static, 2)"):
            for i in range(len(xs)):
                xs[i] = i
    print(xs)   
    \end{lstlisting}
    
    \begin{lstlisting}[language=Python]
from omp4py import *

def f():
    xs = [0] * 20
    def _omp_parallel_1():
        nonlocal xs
        for i in _omp_range(0, len(xs), 1, schedule='static', chunks=2):
            xs[i] = i
            
    _omp_parallel_run(_omp_parallel_1)
            
    print(xs) 
    \end{lstlisting}
	\vspace{-0.35cm}
	\caption{Example of the \texttt{for} directive: user code (top) and its corresponding translation by OMP4Py (bottom).}
	\label{fig:omp_for}    
\end{figure}

Figure \ref{fig:omp_for} shows an example of a loop parallelized with the \texttt{for} directive (top) and its internal implementation generated by OMP4Py (bottom). First, we need to note that the \texttt{for} directive must be within a \texttt{parallel} directive to distribute the work (lines 6-7, top code). Unlike the \texttt{parallel} directive, the \texttt{for} directive must be the unique element within the \texttt{with} block; no other statements are allowed inside the block except within the loop itself. The implementation of this directive is simple: the \texttt{range} function (line 8, top code) is replaced by \texttt{\_omp\_range} (line 7, bottom code), and the scheduling and chunk values are passed as keyword arguments to the function. The \texttt{\_omp\_range} function is always generated with the three positional arguments of \texttt{range}, even if the user does not explicitly specify them. The function, implemented in the OMP4Py runtime, returns a different iterator for each thread, which behaves according to the selected scheduling policy by returning the values assigned to that thread one by one. Once the iterator runs out of elements, it blocks until all other iterators have finished. This synchronization is achieved because all threads in the current team share a barrier object: once an iterator completes, it waits at the barrier until all threads reach it. As previously mentioned, this behavior can be avoided using the \texttt{nowait} clause.

Furthermore, the \texttt{for} directive can be extended to multiple nested loops using the \texttt{collapse} clause. Parallelizing multiple nested loops increases the number of iterations available for distribution among threads. This is especially useful when the outer loops have relatively few iterations but the total number of iterations across all loops is large. When the \texttt{collapse} clause is used, the OMP4Py runtime effectively flattens the nested loops into one loop. The number of nested loops to collapse is specified as an argument to the clause. The loops must be perfectly nested, and the number of iterations in the inner loops must not depend on the iteration variable of an outer loop. Figure \ref{fig:omp_for_collapse} shows an example of using the \texttt{collapse} clause and its implementation process. The translation is similar to the previous case: the ranges of the two loops (lines 9-10, top code) are combined into a single loop where \texttt{\_omp\_range} (line 10, bottom code) receives the arguments of both ranges in tuple format and returns the values for both index variables. Finally, the example also illustrates the use of the \texttt{lastprivate} clause, which is similar to the \texttt{private} clause, but with the added feature that the variable is updated with the value assigned in the last iteration of the loop. A new variable (line 9, bottom code) is created to replace the original variable (line 6, top code) using the \texttt{\_omp\_lastprivate} function (line 14, bottom code), which employs an if statement to determine which thread performed the last iteration and updates the value of the original variable (line 15, bottom code).

\begin{figure}[t!]
    \begin{lstlisting}[language=Python]
from omp4py import *

@omp
def f():
    xs = [0] * 20
    x = 0
    with omp("parallel"):
        with omp("for schedule(static, 2) collapse(2) lastprivate(x)"):
            for i in range(len(xs)):
                for i in range(4):
                    x = j
                    xs[i] += x
    print(x, xs)   
    \end{lstlisting}
    
    \begin{lstlisting}[language=Python]
from omp4py import *

def f():
    xs = [0] * 20
    x = 0
    def _omp_parallel_1():
        nonlocal xs
        nonlocal x
        omp_x = None
        for i, j in _omp_range((0, 0), (len(xs), 4), (1, 1), schedule='static', chunks=2):
            omp_x_1 = j
            xs[i] += omp_x_1

        if _omp_lastprivate(i, j):
            x = omp_x_1
            
    _omp_parallel_run(_omp_parallel_1)
            
    print(x, xs)  
    \end{lstlisting}
	\vspace{-0.35cm}
	\caption{Example of the \texttt{for} directive with \texttt{collapse} and \texttt{lastprivate} clauses: user code (top) and its corresponding translation by OMP4Py (bottom).}
	\label{fig:omp_for_collapse}    
\end{figure}

\subsubsection{Sections}
\label{sec:sections_construct}

The \texttt{sections} directive divides the work among threads by assigning each thread a structured block defined with the \texttt{section} directive. Therefore, only blocks with the \texttt{section} directive can exist within a \texttt{sections} directive. Each block will be executed once by a single thread, but the order of execution is not predetermined, and any thread may execute one or multiple blocks depending on availability. In sequential execution, the blocks are executed in the order of their definition. Similar to the \texttt{for} directive, there is a synchronization barrier that blocks all threads in the current team until all section blocks have been executed. This barrier can also be removed using the \texttt{nowait} clause. 

Figure \ref{fig:omp_sections} shows an example of the \texttt{sections} directive with three section blocks (top) and its implementation by OMP4Py (bottom). Each block performs a print operation to display a number. In sequential execution, the output sequence is always '\texttt{1 2 3}', but with parallel execution, the numbers can be printed in any order. The implementation uses a \texttt{with} block along with the \texttt{\_omp\_sections} function (line 5, bottom code), which serves two purposes: first, it sets each section block as unexecuted, and second, it provides an exit barrier when all blocks have been executed. This is possible because the \texttt{with} block calls the \texttt{\_\_enter\_\_} method on entry and the \texttt{\_\_exit\_\_} method on exit. This mechanism ensures that if \texttt{nowait} is not used, a call to the barrier is made at the end of the block. Additionally, a unique identifier is assigned to each section block. The \texttt{\_omp\_section} function (lines 6, 8 and 10, bottom code) and the \texttt{if} block check whether the block has been executed or needs to be executed. If the block has not yet been executed, the function returns true and allows one thread to execute it. Afterward, any subsequent call to \texttt{\_omp\_section} for that block will return false, preventing additional threads from executing it. 

Finally, although not shown in the example, the sections directive also supports the \texttt{lastprivate} clause, similar to the \texttt{for} directive detailed previously. Its internal implementation mirrors that of the \texttt{for} directive, using the identifier number of the last \texttt{section} block as an argument to \texttt{\_omp\_lastprivate}, which checks if the invoking thread executed that block.

\begin{figure}[t!]
    \begin{lstlisting}[language=Python]
from omp4py import *

@omp
def f():
    with omp("parallel"):
        with omp("sections"):
            with omp("section"):
                print(1)
            with omp("section"):
                print(2)
            with omp("section"):
                print(3)
    \end{lstlisting}
    
    \begin{lstlisting}[language=Python]
from omp4py import *

def f():
    def _omp_parallel_1():
        with _omp_sections():
            if _omp_section(0):
                print(1)
            if _omp_section(1):
                print(2)
            if _omp_section(2):
                 print(3)
            
    _omp_parallel_run(_omp_parallel_1) 
    \end{lstlisting}
	\vspace{-0.35cm}
	\caption{Example of the \texttt{sections} directive: user code (top) and its corresponding translation by OMP4Py (bottom).}
	\label{fig:omp_sections}    
\end{figure}

\subsubsection{Single}
\label{sec:single_construct}

The \texttt{single} directive specifies a block within a parallel region that will be executed only once by a single thread. The thread that executes the block is indeterminate; it will be the first thread to reach the block. In sequential execution, the directive has no effect since there are no additional threads to execute the block. As with other worksharing directives, the remaining threads must wait until the block's execution is complete before proceeding. This barrier can also be removed using the \texttt{nowait} clause. 

\begin{figure}[t!]
    \begin{lstlisting}[language=Python]
from omp4py import *

@omp
def f():
    x = 0
    with omp("parallel firstprivate(x)"):
        with omp("single copyprivate(x)"):
            x += 1
        print(x)
    \end{lstlisting}
    
    \begin{lstlisting}[language=Python]
from omp4py import *

def f():
    x = 0
    def _omp_parallel_1():
        _omp_x_1 = x
        with _omp_single() as _omp_1:
            if _omp_1:
                _omp_x_1 += 1
                _omp_copyprivate_set(_omp_x_1)
        _omp_x_1 = _omp_copyprivate_get()
            
    _omp_parallel_run(_omp_parallel_1) 
    \end{lstlisting}
	\vspace{-0.35cm}
	\caption{Example of the \texttt{single} directive: user code (top) and its corresponding translation by OMP4Py (bottom).}
	\label{fig:omp_single}    
\end{figure}

Figure \ref{fig:omp_single} shows an example of user code when using a \texttt{single} directive (top) and its internal implementation by OMP4Py (bottom). In this code, a variable declared with the \texttt{firstprivate} clause (line 6, top code) creates a private copy for each thread, initialized with the value from before the parallel region begins. Inside the \texttt{single} block, the variable is incremented by one (line 8, top code). After the \texttt{single} block finishes, the value of the variable from the thread that executed the block is broadcast to all other threads in the team using the \texttt{copyprivate} clause. Note that variables listed in the \texttt{copyprivate} clause must be private to each thread. Once the broadcast is complete, threads in the team have the updated value and print it (line 9, top code).

The implementation of the \texttt{single} directive is similar to that of the \texttt{sections} directive, requiring a \texttt{with} block to control thread entry and exit. The \texttt{\_omp\_single} function (line 7, bottom code) returns true only for the first thread to invoke it, returning false for all other threads. This return value is stored in a temporary variable, and the \texttt{if} block ensures that only the thread that received true executes the directive's code (line 8, bottom code). The \texttt{copyprivate} clause adds a call to the \texttt{\_omp\_copyprivate\_set} function at the end of the block (line 10, bottom code), which takes all variables listed in the clause as arguments. These values are stored internally and then updated in other threads using \texttt{\_omp\_copyprivate\_get} (line 11, bottom code) through multiple assignments. The synchronization barrier in the \texttt{with} block ensures that no thread calls \texttt{\_omp\_copyprivate\_get} before \texttt{\_omp\_copyprivate\_set} has been invoked by the thread that executed the directive. Removing this barrier with \texttt{nowait} would require a more complex, inefficient, and hard-to-debug mechanism. Consequently, the OpenMP standard restricts the use of \texttt{copyprivate} and \texttt{nowait} together in the same \texttt{single} directive.

\subsection{Tasking Directives}
\label{sec:task_construct}

OpenMP tasking provides a flexible way to parallelize heterogeneous or dynamic workloads. The \texttt{task} directive allows for the creation of tasks, which are units of work that can be executed by any thread in the team. This is particularly useful for applications with recursive algorithms, irregular loops, or other patterns where the workload cannot be distributed uniformly across threads from the beginning. When a thread reaches a task directive, it wraps the associated block of code and its data environment into a task, which can then be executed by any thread in the current team. A task does not need to be executed immediately, it can be stored in a queue for later execution. Threads in the team can dynamically take tasks from the queue and execute them. Tasks can be executed explicitly with the \texttt{taskwait} directive or implicitly when a thread becomes available. In any case, all tasks will be completed before the thread team terminates. 

The implementation of the \texttt{task} directive is similar to \texttt{parallel} in the sense that \texttt{parallel} can be considered as a task executed by all threads in the team at its start. The definition of the variable scope for creating the environment follows the same principles, utilizing clauses such as \texttt{default}, \texttt{private}, \texttt{firstprivate}, and \texttt{shared}. There is an additional \texttt{if} clause that allows for conditional task creation: if the condition evaluates to false, the task is executed immediately instead of being enqueued.

\begin{figure}[t!]
    \begin{lstlisting}[language=Python]
from omp4py import *

@omp
def fib(n):
    i = 0
    j = 0
    if n < 2:
        return n
    with omp("task"):
        i = fib(n - 1)
    with omp("task"):
        j = fib(n - 2)
    omp("taskwait")
    return i + j

@omp
def f(n):
    x = 0
    with omp("parallel"):
        with omp("single"):
            x = fib(n)
    print(x)
    \end{lstlisting}
    
    \begin{lstlisting}[language=Python]
from omp4py import *

def fib(n):
    i = 0
    j = 0
    if n < 2:
        return n
    def _omp_task_1():
        nonlocal i
        nonlocal j
        i = fib(n - 1)
    _omp_task_submit(_omp_task_1)
    def _omp_task_2():
        nonlocal i
        nonlocal j
        j = fib(n - 2)
    _omp_task_submit(_omp_task_2)
    _omp_taskwait()
    return i + j

def f():
    x = 0
    def _omp_parallel_1():
        nonlocal x
        with _omp_single() as _omp_1:
            if _omp_1:
                x = fib(n)
            
    _omp_parallel_run(_omp_parallel_1) 
    \end{lstlisting}
	\vspace{-0.35cm}
	\caption{Fibonacci calculation example using the tasking directives: user code (top) and its corresponding translation by OMP4Py (bottom).}
	\label{fig:omp_task}    
\end{figure}

Figure \ref{fig:omp_task} shows the recursive calculation of Fibonacci numbers using tasks (top) and its internal implementation generated by OMP4Py (bottom). The algorithm defines the Fibonacci function (line 4, top code), which calculates the value for a given $n$ by summing the recursive calls to the same function for the values $n-1$ and $n-2$ (lines 10 and 12, top code). The initial call to the Fibonacci function is placed in a single block to ensure it is executed by only one thread (line 21, top code), while the remainder threads are blocked. The function then creates a task for the recursive calculation (lines 9 and 11, top code) and blocks the current task until they have been executed (line 13, top code). The threads in the team will consume the created tasks and generate new tasks until they reach the base cases of $n=0$ and $n=1$. The code generated by OMP4Py shows the similarities with the \texttt{parallel} directive. In this way, the code within the \texttt{task} directive (lines 9 and 11, top code) and the \texttt{parallel} directive (line 19, top code) both result in the creation of functions: \texttt{\_omp\_task\_1}, \texttt{\_omp\_task\_2} (lines 8 and 13 in the bottom code), and \texttt{\_omp\_parallel\_1} (line 23, bottom code), respectively. The only notable difference is that \texttt{parallel} uses the function \texttt{\_omp\_parallel\_run} (line 29, bottom code), while task employs \texttt{\_omp\_task\_submit} (lines 12 and 17, bottom code) which places the task function into a shared queue. The queue is accessible by all threads in the team that will be able to pick up the task and execute it. Finally, the \texttt{taskwait} clause is implemented as a simple function (line 18, bottom code), which forces the caller thread to consume all tasks in the queue until it is empty, at which point the function returns. 

\subsection{Interaction with Python}

In this section, we will explain the implementation of the \texttt{\_omp} internal OMP4Py API functions that were introduced in the previous sections and how they interact with Python. The generated code uses the internal API to manage operations such as thread management and work distribution, all implemented with Python code and standard libraries, making OMP4Py a pure native Python library. 

When a Python program using OMP4Py is executed, a single thread begins by running the main program, as in a typical Python script. However, it is important to note that before the main program starts, the Python interpreter processes all decorators, including \texttt{omp} decorators. So, by the time the main program starts, all OpenMP directives within the source code will be replaced with a new pure Python version with parallel behavior. According to the OpenMP standard, this thread is called the initial thread. OMP4Py initializes the initial thread context with the first call to any function in its API. The remaining threads created in parallel constructs will be initialized with a context derived from the initial thread before they begin execution. For this reason, any thread created outside the OMP4Py clauses, for instance, using \texttt{asyncio}, \texttt{concurrent.futures}, \texttt{threading}, or the \texttt{multiprocessing} module, will not have an associated context and will be treated by OMP4Py as an uninitialized initial thread. If there is a subsequent call to any function in the OMP4Py API by one of these threads, a new OMP4Py context will be created for it. In this way, the thread will act as a new initial thread and can independently use all the features of OMP4Py, leaving the user responsible for managing potential concurrency issues between different instances.

The context in OMP4Py is implemented by a stack object where OpenMP construct tasks are stored. In the OpenMP standard, the initial thread exists within a single-threaded parallel region. Therefore, the initialization of the initial thread's context involves adding a \texttt{parallel} construct task to the stack, allowing the OpenMP API (like \texttt{omp\_get\_thread\_num}) to be called from any point in the source code, ensuring consistent results. The context is stored locally to the thread using \texttt{threading.local}.

In a typical execution flow of an OpenMP application, the next task added to the stack is the \texttt{parallel} task, which creates a parallel region with multiple threads. The \texttt{\_omp\_parallel\_run} function is responsible for this process. It works as follows: the initial thread creates a context for each new thread, with a stack that is a shadow copy of its own. The initial thread then adds the \texttt{parallel} task to all stacks, including its own, and finally, using the threading library, the threads are created with their respective contexts as arguments. The \texttt{parallel} task contains the following information: the thread ID, the number of threads, a mutex, a barrier, a shared task list, and a shared dictionary. The thread ID and number of threads are critical for work distribution and API functions. The mutex and barrier are used for constructs like \texttt{critical} and \texttt{barrier}. The shared task list holds tasks (created by the \texttt{task} construct) that must be completed before the parallel region ends, while the shared dictionary stores common data, such as iteration counters for dynamic workloads in a \texttt{for} construct. Exceptions thrown within a parallel region are not automatically propagated to the initial thread. Instead, the behavior depends on how the exception is handled within each thread: if an exception occurs, it must be caught and handled within that thread. OMP4Py catches unhandled exceptions in parallel regions to prevent the program from terminating unexpectedly, but relying on this behavior is considered bad practice.

As explained in Section \ref{sec:for_construct}, the \texttt{for} construct is executed by each thread in the parallel region using the \texttt{\_omp\_range} function, which returns an iterator for the thread's assigned values. For static scheduling, each thread divides the work based on its thread ID and returns an independent iterator. For dynamic or guided scheduling, the thread acquires the mutex, checks the shared dictionary for an iteration counter (creating it if needed), and uses it to fetch new work chunks. This is done using Python iterators and the \texttt{yield} function. The \texttt{sections} and \texttt{section} constructs (Section \ref{sec:sections_construct}) work similarly, where \texttt{\_omp\_sections} creates a sections task with an empty set, and each \texttt{\_omp\_section} function checks and updates the set to track executed sections, all while using the mutex to avoid race conditions.

The \texttt{\_omp\_single} function (see Section \ref{sec:single_construct}) ensures that only one thread within a parallel region executes a section of code. It does so by returning the mutex of the region, which locks the code block for exclusive access by one thread using the \texttt{with} statement. This causes the mutex to be locked at the start of the block and released automatically when the block finishes. 

The functions \texttt{\_omp\_copyprivate\_get} and \texttt{\_omp\_copyprivate\_set} facilitate data sharing between threads. \texttt{\_omp\_copyprivate\_get} retrieves variables from the shared dictionary and makes threads wait for data using \texttt{threading.Condition}, while \texttt{\_omp\_copyprivate\_set} stores data and notifies other threads when it becomes available.

Finally, the \texttt{\_omp\_task\_submit} function (Section \ref{sec:task_construct}) adds a task to the shared task list of the current parallel region, enabling it to be queued for execution. Tasks are then assigned to available threads for execution, either when the parallel block finishes or through the \texttt{taskwait} construct. The \texttt{\_omp\_taskwait} function ensures that a thread consumes tasks from the shared list, only returning when the list is empty. To prevent race conditions, a mutex is used each time a thread removes a task for execution.

\section{Experimental Results}
\label{sec:exp_results}

Next, we will evaluate the performance and scalability of different Python applications parallelized using OMP4Py. Additionally, we will highlight the differences and advantages of having a native OpenMP implementation in Python, as provided by OMP4Py, compared to PyOMP, which is a Numba-based prototype with OpenMP support.

Experiments were conducted using one server with two 24-core Intel Xeon Gold 5220R @2.2GHz processors and 192 GB of RAM. The software used was Python v3.13, NumPy v2.1.1, mpi4py v4.0.0, NetworkX v3.3, and PyOMP v0.1 (September 2024). Execution times were averaged over 10 measurements for each test.

\begin{figure*}[t]
	\centering
	\includegraphics[width=0.65\textwidth]{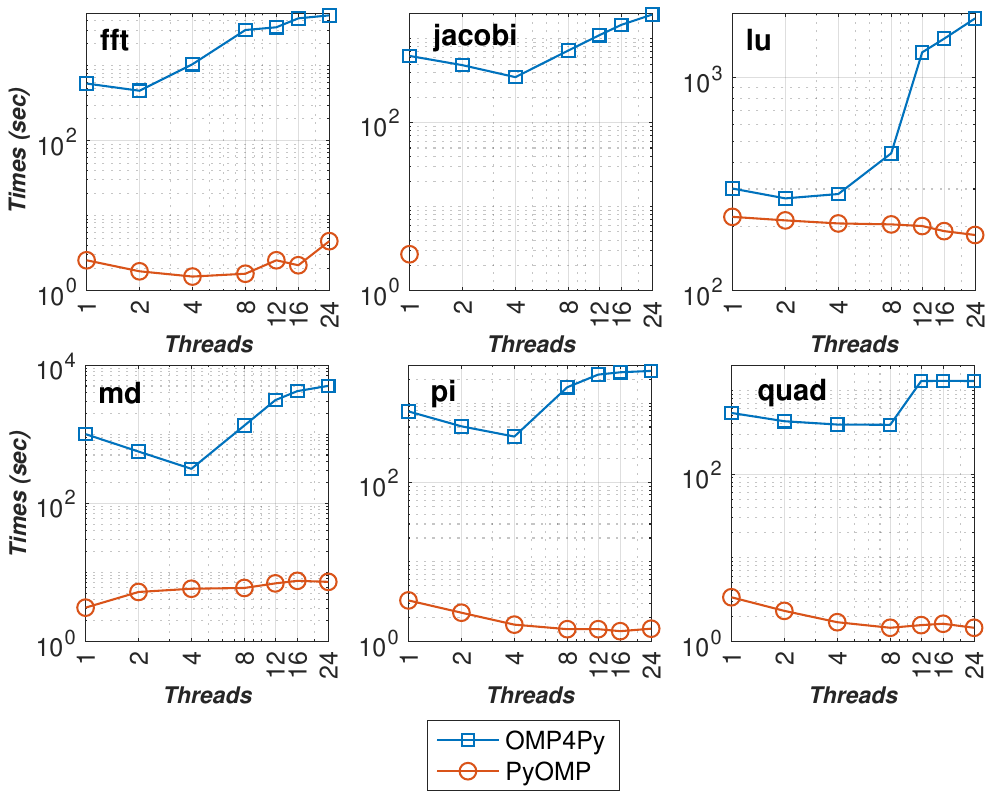}
	\caption{Scalability of the different parallel numerical applications. Axis in log scale.}
	\label{fig:num_alg_times} 
\end{figure*}

\subsection{Numerical algorithms}
\label{sec:perf_num_algorithms}

We have selected six algorithms that represent different types of numerical application patterns to evaluate the performance and scalability of OMP4Py. In particular: 
\begin{itemize}
    \item \emph{Fast Fourier Transform (fft)}. It is an efficient algorithm used to compute the Discrete Fourier Transform (DFT) of a sequence, allowing for the conversion of a signal from its time domain to its frequency domain. Performance tests were run using a complex data vector of 4 million numbers.
    
    \item \emph{Jacobi method (jacobi)}. It is an iterative algorithm for solving systems of linear equations of the form \(A \cdot x = b\), where \(A\) is a matrix, and \(x\) and \(b\) are vectors. At each iteration, the solution vector is updated based only on values from the previous iteration. We used a square matrix \(A\) of size \(1k \times 1k\), performing up to 1,000 iterations, with a stopping criterion of an error tolerance of \(1 \times 10^{-6}\).
    
    \item \emph{LU decomposition (lu)}. It is a method used to factor a matrix \(A\) into the product of two matrices: a lower triangular matrix \(L\) and an upper triangular matrix \(U\), such that \(A = L \cdot U\). This factorization simplifies solving systems of linear equations, matrix inversion, and determinant computation. We applied LU decomposition to a square matrix of size \(1k \times 1k\).
    
    \item \emph{Molecular dynamics simulation (md)}. The simulation was conducted to study the motion of particles over time. The velocity Verlet integration scheme was employed to update positions, velocities, and accelerations. We simulated a system of 2,000 particles interacting with a central pair potential.
    
    \item \emph{Computing $\pi$ (pi)}. The area under the curve \(y = \frac{4}{1 + x^2}\) between 0 and 1 provides an approximation for \(\pi\). This integral can be estimated using numerical summation, where we used 2 billions of intervals to compute the approximation.

    \item \emph{QUAD}. It is a numerical integration technique that estimates the value of an integral using an averaging method. It approximates the integral of the function \( f(x) = \frac{50}{\pi \cdot (2500 \cdot x^2 + 1)} \) over the interval from \(A = 0\) to \(B = 10\). This method involves sampling the function at numerous points within the interval to compute an average value, which is then used to estimate the integral. For our tests, we employed 1 billion iterations.
\end{itemize}

All the source codes can be found in the OMP4Py repository. Figure \ref{fig:num_alg_times} shows the execution times of the different parallel benchmarks using from 1 to 24 threads. It can be observed that the scalability of the OMP4Py applications is not good. For instance, only 4 out of 6 benchmarks benefit from using 4 threads, while using more threads consistently proves detrimental to performance. The best overall speedup, calculated as the ratio of sequential time to parallel time, is 3.18$\times$, obtained for the molecular dynamics application using 4 threads. 

\begin{figure}[t]
	\centering
	\includegraphics[width=0.75\linewidth]{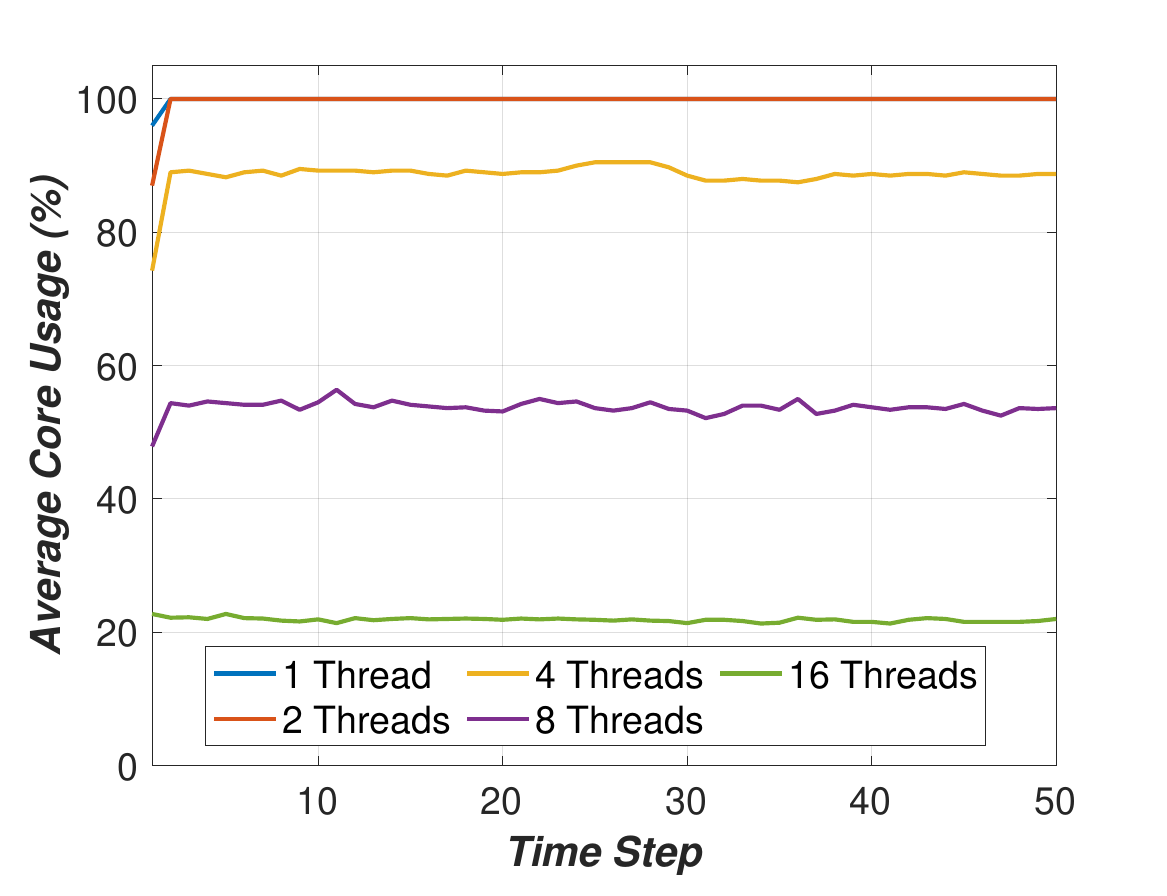}
	\caption{Average core usage when running the \emph{pi} application using different number of threads. Time step is 5 seconds.}
	\label{fig:cpu_pi_versions} 
\end{figure}

To better understand the behavior of OMP4Py parallel applications, as an illustrative example, we will focus on tracking core usage while executing the \emph{pi} application with different thread counts (Figure \ref{fig:cpu_pi_versions}). The profiling was performed using the \emph{mpstat} command, which is part of the \emph{sysstat} package. This tool provides detailed statistics on CPU usage, allowing us to monitor individual core usage and overall system performance. The data was captured at 5-second intervals. As observed, only with 1, 2, and 4 threads does the average core usage reach 100\% or close to it, which corresponds to the cases where the parallel code scales efficiently (see Figure \ref{fig:num_alg_times}). However, a significant decrease in core usage is detected as the number of threads increases. Specifically, average values of only about 54\% and 22\% were obtained when using 8 and 16 threads, respectively. The fact that the average core usage remains low suggests that thread synchronization overhead (e.g., waiting for shared resources) likely prevents the parallel application from scaling effectively. Note that this behavior was already identified by Python developers and currently remains an open issue~\footnote{\url{https://github.com/python/cpython/issues/118649} [online, accessed February 27, 2025]}. In any case, as we will demonstrate next through several experiments, it is not related to the Python code generated by OMP4Py.

First, we aim to estimate the baseline overhead introduced by OMP4Py. To do so, we measured the execution times of numerical applications running with OMP4Py using a single thread and in a purely sequential manner, without any explicit reference to OMP4Py in the user code. Ten measurements were carried out for each version and application. The execution time differences between the OMP4Py (single-thread) and sequential implementations vary slightly across applications but remain consistently low, with all differences below 0.2\%. The results of the Wilcoxon rank-sum test indicate that these differences are not statistically significant in all cases, meaning OMP4Py does not introduce a meaningful impact on execution time.

Once we established that the overhead introduced by OMP4Py is negligible, we will investigate the origin of the poor scalability by implementing, as a representative case, three different versions of the \emph{pi} application. The user codes are shown in Figure \ref{fig:omp_pi_versions}. The first approach (top code) is a version in which thread creation, loop iteration distribution across threads, and final reduction of the $\pi$ value are all handled automatically by OMP4Py. The second approach (middle code) replaces the automatic distribution of loop iterations performed by OMP4Py with manual distribution. While thread creation and the final reduction are still managed by OMP4Py, the code explicitly calculates the start and end indices for each thread. The \texttt{compute} function is called (lines 1–5, middle code), taking the calculated range and iterating through it to perform the $\pi$ calculation. The workload division is based on the number of threads, \texttt{k}, and each thread's identifier, \texttt{tid}, which are retrieved using the \texttt{omp4py.omp\_get\_num\_threads()} and \texttt{omp4py.omp\_get\_thread\_num()} functions (lines 12–13, middle code). This approach represents the explicit version of the code that OMP4Py would have generated automatically in the first approach. The third version (bottom code) builds upon the second version but incorporates Cython~\cite{Beh10} to optimize the \texttt{compute} routine. Cython compiles Python code into highly optimized C code, which is then compiled into a shared library. This allows Python to call the function as a native extension, bypassing the Python interpreter and significantly improving performance. By applying the \texttt{@cython.compile} decorator (line 1, bottom code), the computation of $\pi$ is further accelerated, as the \texttt{compute} function is compiled into efficient C code. Note that thread creation and reduction for computing $\pi$ remain automatically managed by OMP4Py, while loop iteration distribution remains manual. This is because the \texttt{omp4py\_pi} function is identical to the second version but with Cython’s performance enhancements.
    
\begin{figure}[t!]
\centering
\begin{minipage}{0.45\textwidth}
\begin{lstlisting}[language=Python]
@omp
def omp4py_pi(n):
    w = 1.0 / n
    PI = 0.0
    with omp("parallel for reduction(+:PI)"):
        for i in range(n):
            local = (i + 0.5) * w
            PI += 4.0 / (1.0 + local * local)
    return PI * w
\end{lstlisting}
\end{minipage}  
\begin{minipage}{0.45\textwidth}
\begin{lstlisting}[language=Python]
def compute(start, end, w, PI):
    for i in range(start, end):
        local = (i + 0.5) * w
        PI += 4.0 / (1.0 + local * local)
    return PI

@omp
def omp4py_pi(n):
    w = 1.0 / n
    PI = 0.0
    with omp("parallel reduction(+:PI)"):
        k = omp4py.omp_get_num_threads() 
        tid = omp4py.omp_get_thread_num()
        chunk = n // k
        rem = n % k

        start = chunk * tid
        end = start + chunk + (1 if tid < rem else 0)
        PI = compute(start, end, w, PI)
    return PI * w
\end{lstlisting}
\end{minipage}   
\begin{minipage}{0.45\textwidth}
\begin{lstlisting}[language=Python]
@cython.compile
def compute(start: cython.long, end: cython.long, 
            w:cython.float, PI:cython.float) -> cython.float:
    i: cython.long
    for i in range(start, end):
        local:cython.float = (i + 0.5) * w
        PI += 4.0 / (1.0 + local * local)
    return PI
\end{lstlisting}
\end{minipage}    
\vspace{-0.3cm}
\caption{User codes for the \emph{pi} application: (top) OMP4Py (middle) manually parallelized, and (bottom) \texttt{compute} function using Cython.}
\label{fig:omp_pi_versions}    
\end{figure}

\begin{figure}[t]
	\centering
	\includegraphics[width=0.75\linewidth]{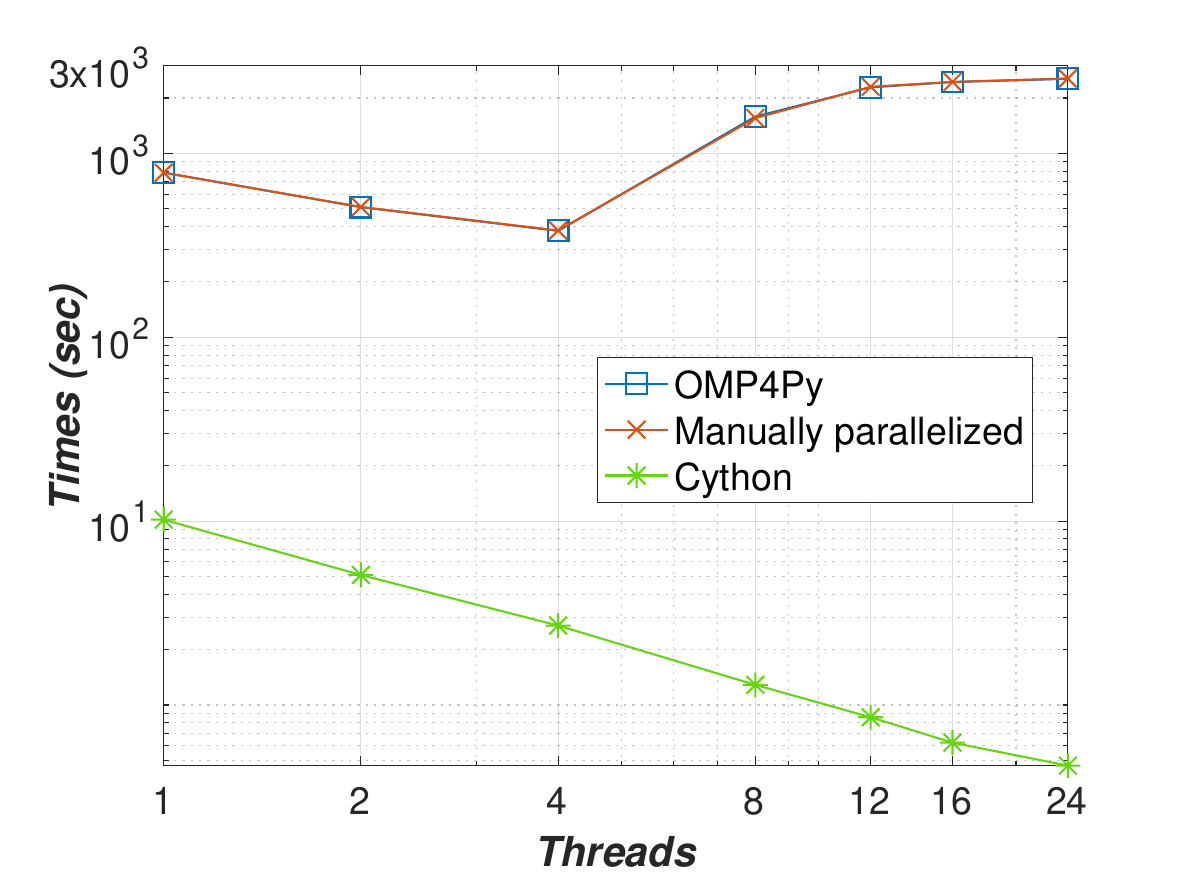}
	\caption{Scalability of the different user codes shown in Figure \ref{fig:omp_pi_versions} for the \emph{pi} application. Axis in log scale.}
	\label{fig:times_pi_versions} 
\end{figure}

Figure \ref{fig:times_pi_versions} presents the execution times of the three code versions of the \emph{pi} benchmark, using 1 to 24 threads. Note that the results for the OMP4Py version (top code in Figure \ref{fig:omp_pi_versions}) are the same as those displayed in Figure \ref{fig:num_alg_times} for the \emph{pi} application. From these results, several conclusions can be drawn:
\begin{itemize}
    \item Since the performance of the OMP4Py version and the manually parallelized version are identical, it can be concluded that the overhead of the automatic loop iteration distribution performed by OMP4Py is negligible. Furthermore, scalability issues persist in both versions, indicating that they are not caused by work distribution. 
    \item As expected, the execution times when using Cython to compile the \texttt{compute} function are significantly faster than when using pure Python. We must highlight that, with one thread, pure Python takes about 785 seconds to calculate $\pi$, whereas the Cython version requires only about 10 seconds. Since the \emph{pi} application is a compute-intensive benchmark, this behavior confirms that these performance differences are solely due to the fundamental differences between compiled (C) and interpreted (Python) languages.  
    \item The scalability of the Cython code is excellent; there are no signs of scalability issues. For example, the speedup reaches 21.7$\times$ when using 24 threads. Considering that the only difference between the manually parallelized version (middle code in Figure \ref{fig:omp_pi_versions}) and the Cython version is the \texttt{compute} routine, we can conclude that the scalability problems are not related to OMP4Py, since the overhead of thread creation, loop iteration distribution, and final reduction remains the same. Therefore, the only plausible explanation is that inefficiencies exist in the Python interpreter when multiple threads execute the \texttt{compute} routine concurrently. In any case, these issues cannot be attributed to the code generated by OMP4Py.
\end{itemize}

The previous analysis demonstrates that the current version of the Python interpreter (v3.13) still lacks mature support for multithreading, with several unresolved implementation issues that hinder its efficient use. In particular, as of February 2025, Python developers are dealing with over 100 open issues labeled under the free-threading topic—half of which are bugs, and 15 of which correspond to critical crashes. While this new version is an important step forward in removing the GIL, it still limits the use of OMP4Py to a small number of threads when running numerical (compute-intensive) applications. However, it is important to highlight that as these issues in the Python interpreter are progressively resolved, the scalability limitations of OMP4Py when running numerical applications will gradually disappear. On the other hand, as we will show in Section \ref{sec:app_objects}, non-numerical OMP4Py applications exhibit strong scalability with codes containing the same structure and OpenMP directives as those discussed in this section.

Finally, for illustrative purposes, Figure \ref{fig:num_alg_times} also shows the performance results obtained by the applications when compiled using PyOMP (Numba). As expected, they are significantly faster than pure Python, as Numba compiles Python code into optimized machine code, much like Cython, bypassing the slower interpreted execution of standard Python. This leads to substantial reductions in the execution times, particularly for compute-intensive tasks such as numerical computations. In any case, we have observed problems in the scalability of some benchmarks. For example, with 24 threads, the best speedups were obtained for the \emph{quad} and \emph{pi} applications. However, these speedups are quite limited, reaching only 2.32$\times$ and 2.27$\times$, respectively. This corresponds to a low parallel efficiency of just 9.7\% and 9.5\%, calculated as the ratio of speedup to the number of threads used, expressed as a percentage. \emph{fft} scales only up to 4 threads, while performance degrades for higher thread counts. On the other hand, the \emph{lu} and \emph{md} applications do not scale at all, and their execution times increase as the number of running threads grows. Additionally, PyOMP was unable to execute the \emph{jacobi} method using more than one thread.

\subsection{Applications using non-numerical libraries and Python objects}
\label{sec:app_objects}

As we pointed out previously, PyOMP is a fork of the Numba project. The Numba library's \texttt{njit} decorator accelerates Python functions by compiling them to machine code, but it imposes limitations on using features such as functions from non-Numba-optimized libraries and certain Python objects and data structures. Numba provides special support for some commonly used libraries like math and numpy, allowing their functions to be called within \texttt{@njit}-compiled functions, as these libraries are optimized for machine code compilation. As a result, Numba restricts function usage to those either decorated with \texttt{@njit} or from libraries that Numba specifically optimizes. In contrast, OMP4Py, being a native OpenMP implementation for Python, is designed to work with a broader range of code, including those that may not be compatible with Numba's restrictions. 

To illustrate the benefits of having this native OpenMP implementation for Python, we have implemented the following applications using OMP4Py:

\begin{itemize}
    \item \emph{Graph Clustering}. The clustering coefficient of a node is the fraction of possible triangles that pass through that node in an unweighted graph. We used a graph with 300k vertices, each connected by 100 edges, as input. The graph generation, storage, and clustering algorithm were implemented using the NetworkX~\cite{Hag08} library. Note that PyOMP cannot run this benchmark because Numba is unable to compile the object \emph{Graph} and the clustering algorithm function calls, as they are part of an external library that is not optimized for Numba.
    
    \item \emph{Wordcount}. It is a simple algorithm that counts the number of occurrences of each word in an input text. We generated a text consisting of 1 million characters, featuring words with lengths between 3 and 10 letters, with a 10\% probability that a new line will be added after each word. Note that although more recent versions of Numba have increasingly added experimental support for Python dictionaries, PyOMP is a fork of an earlier version that lacks the necessary support to compile Wordcount dictionaries.
\end{itemize}

\noindent Source codes can be found in the OMP4Py repository.

\begin{figure}[t]
	\centering
	\includegraphics[width=\linewidth]{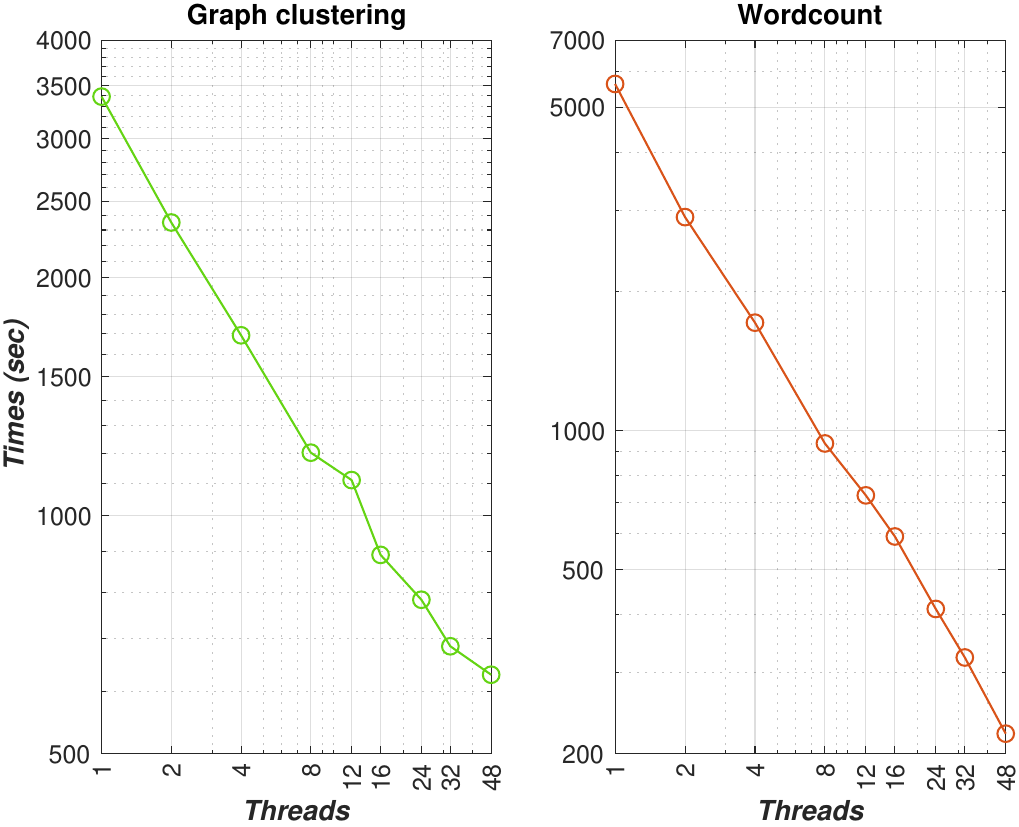}
	\caption{Scalability of the Graph Clustering (left) and Wordcount (right) applications. Axis in log scale.}
	\label{fig:ext_lib_times} 
\end{figure}

Figure \ref{fig:ext_lib_times} shows the scalability of the Graph Clustering and Wordcount applications using up to 48 threads. The scalability of OMP4Py is good, especially for Wordcount, which achieves a 25.5$\times$ speedup compared to the sequential execution when using 48 threads. This behavior is very different from the performance results obtained for the numerical applications (see Figure \ref{fig:num_alg_times}). However, the structure and OpenMP directives used in both types of applications are very similar, with the only differences being related to the operations performed inside the loops. For example, when closely examining the OMP4Py codes for the $\pi$ calculation (Figure \ref{fig:omp_pi_versions}, top code) and the Wordcount application (Figure \ref{fig:omp_wordcount}), it can be observed that the OpenMP clauses are semantically identical. In the case of a dictionary reduction, the \texttt{reduction} clause cannot be used, and manual implementation with a \texttt{critical} block is necessary. On the other hand, for numerical variables, the reduction clause can be used, and OMP4Py automatically generates the same corresponding \texttt{critical} block. This additional observation, together with the analysis in Section \ref{sec:perf_num_algorithms}, confirms that the scalability issues detected in the $\pi$ calculation cannot be attributed to OMP4Py, as the generated code for both applications is essentially identical. Therefore, these results can only be explained by the fact that free-threaded support for Python 3.13 is still in an experimental stage, leaving plenty of room for improving performance and efficiency.

\begin{figure}[t!]
\centering
\begin{minipage}{0.45\textwidth}
\begin{lstlisting}[language=Python]
from omp4py import *

@omp
def wordcount(lines):
    count = {}
    with omp("parallel"):
        local_count = {}
        with omp("for"):
            for i in range(len(lines)):
                for word in lines.split():
                    if word in local_count:
                        local_count[word] += 1
                    else:
                        local_count[word] = 1
        with omp("critical"):
            count.update(local_count)

    return count
\end{lstlisting}
\end{minipage}
\vspace{-0.3cm}
\caption{OMP4Py user code for the Wordcount application.}
\label{fig:omp_wordcount}    
\end{figure}


\subsection{Hybrid applications combining OMP4Py with mpi4py}

One of the main features of OMP4Py is that it can be combined with mpi4py~\cite{dalcin21} to implement hybrid parallel applications that can exploit intra- and inter-node parallelism. The mpi4py package provides Python bindings for the Message Passing Interface (MPI) standard~\cite{mpi41}, which is the most widely used and dominant programming model in HPC.

It is important to highlight that although mpi4py is a Python wrapper for the MPI C library, Numba cannot use MPI code within its functions because it treats mpi4py as an external library. Numba's compilation process focuses on translating Python code to machine code but does not include functionality for integrating or compiling external libraries like MPI automatically. Even though mpi4py interfaces with C, Numba is unaware of how to compile MPI operations into their C equivalents. For this reason PyOMP cannot be combined with mpi4py.

As a case study, we implemented a hybrid parallel version of the Jacobi application described in Section \ref{sec:perf_num_algorithms}. To implement the Jacobi method to solve \(Ax = b\) using MPI, the matrix \(A\) and vector \(b\) are distributed across multiple processors, with each processor managing a portion of the matrix rows and corresponding elements of the vector. During each iteration, processors compute updated values of \(x\) using OpenMP based on their assigned subset of \(A\) and \(b\). MPI function \texttt{MPI\_Allgather} is employed to exchange the updated vector \(x\) among all processors to ensure consistency. Convergence is assessed by computing the global error, with \texttt{MPI\_Allreduce} used to aggregate and verify whether the stopping criterion is satisfied. This code can also be found in the OMP4Py repository.

\begin{figure}[t]
	\centering
	\includegraphics[width=0.8\linewidth]{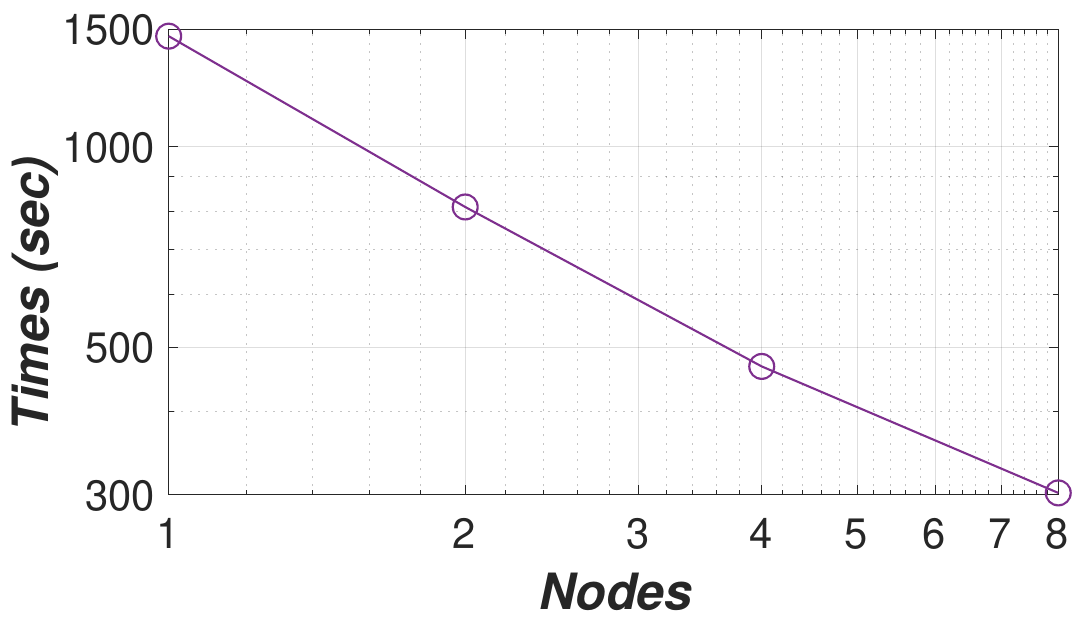}
	\caption{Scalability of a hybrid implementation (OMP4Py + mpi4py) of the Jacobi application using different number of computing nodes. 16 threads per computing node were used. Axis in log scale.}
	\label{fig:jacobi_hyb_times} 
\end{figure}

Experiments in this section were performed on a cluster using up to 8 computing nodes, each node containing two 24-core Intel Xeon Gold 5220R @2.2GHz processors and 192 GB of RAM. The performance results, illustrated in Figure \ref{fig:jacobi_hyb_times}, show speedups of 1.84$\times$, 3.13$\times$ and 4.85$\times$ when using 2, 4, and 8 nodes respectively, compared to using a single node. These strong scalability results demonstrate the potential of OMP4Py to enhance the performance of applications currently parallelized using only mpi4py. Tests were conducted using 16 threads per node.

\section{Conclusions}
\label{sec:conclusions}

In this paper, we introduce OMP4Py\footnote{It is publicly available at \url{https://github.com/citiususc/omp4py}}, a novel implementation of the OpenMP standard designed specifically for Python. It fully supports the API specifications of version 3.0, ensuring that we address most of the needs of OpenMP HPC applications and programmers. OMP4Py integrates OpenMP into Python by adapting its directive-based approach, allowing users to embed parallel constructs directly into their code. These transformer directives enable parallel execution by instructing OMP4Py to modify the Python code as needed.

The performance evaluation of OMP4Py reveals limited scalability when using numerical algorithms. Specifically, only 4 out of 6 benchmarks showed improvement with 4 threads, with performance typically degrading when more threads were used. However, we demonstrated that these scalability issues are due to Python’s multithreading limitations in version 3.13, rather than OMP4Py’s implementation. It is important to highlight that as these limitations in the Python interpreter are progressively resolved, the scalability constraints of OMP4Py when running numerical applications will gradually disappear. 

Non-numerical applications using Python objects and external libraries show much better scalability, achieving up to 25.5$\times$ speedup with 48 threads. This confirms that OMP4Py works well for a broader range of applications compared to PyOMP (Numba), which is restricted to numerical algorithms by its inability to handle these types of data structures and libraries. 

Finally, OMP4Py can be effectively combined with mpi4py to create hybrid parallel applications that leverage both intra- and inter-node parallelism, demonstrating strong scalability. 

In the future, OMP4Py will be extended to support newer versions of the OpenMP standard, including versions 4.0 through 6.0. This will involve incorporating advanced features such as task dependencies, thread teams and support for accelerators. Additionally, the current implementation will be optimized by reducing mutex locks and exploring atomic operations, which are commonly used in C-based OpenMP implementations.

\section*{Declaration of Generative AI and AI-assisted technologies in the writing process}
During the preparation of this work the authors just used
chatGPT-3.5 in order to improve readability. After using this tool/service, the authors reviewed and edited the content as needed and take full responsibility for the content of the publication.

 \bibliographystyle{elsarticle-num} 


\end{document}